\documentclass{article} % For LaTeX2e
\usepackage{iclr2027_conference,times}
\usepackage{graphicx}
\usepackage{sidecap}
\usepackage{amsmath,amsfonts,bm}

\def\eqref#1{equation~\ref{#1}}
\def\1{\bm{1}}

\DeclareMathAlphabet{\mathsfit}{\encodingdefault}{\sfdefault}{m}{sl}
\SetMathAlphabet{\mathsfit}{bold}{\encodingdefault}{\sfdefault}{bx}{n}

\usepackage{hyperref}
\hypersetup{hidelinks}

\usepackage{url}

\title{Alignment of visual brain encoding models with individual perceptual similarity judgments}

\author{
\noindent
\textbf{Daniel Chong$^{1}$, Haomiao Chen$^{1,2}$,
Keith Jamison$^{3}$, Raja Marjieh$^{4}$,Ilia Sucholutsky$^{5}$,} \\
\textbf{Nori Jacoby$^{1}$ \& Amy Kuceyeski$^{1,3}$} \\
$^{1}$Cornell University,
$^{2}$Cornell Tech,
$^{3}$Weill Cornell Medicine, \\
$^{4}$Harvard University,
$^{5}$Purdue University \\
\texttt{djc473@cornell.edu, hc872@cornell.edu, kwj2001@med.cornell.edu,} \\
\texttt{rmarjieh@g.harvard.edu, isucholu@purdue.edu, kj338@cornell.edu,} \\
\texttt{amk2012@med.cornell.edu}
}

\iclrfinalcopy % Uncomment for camera-ready version, but NOT for submission.
\begin{document}

\maketitle

\begin{abstract}
Encoding models that predict brain activation responses to various stimuli, including images, have traditionally been trained to maximize accuracy when predicting the observed brain data. However, in doing so, their internal representations may not reflect individual human perception. For example, two images that have very different predicted brain responses may be perceptually similar to a specific person. In this work, we investigate whether we can fine-tune brain encoding models to better align their representations with specific individuals' similarity judgments. This improved perceptual alignment may be useful in downstream tasks using the brain encoding model, particularly decoding of seen or imagined images. Using measured similarity judgments of 100 images from the 8 participants in the Natural Scenes Dataset (NSD), we adapt a generic human similarity judgment model (DreamSim) separately for each NSD participant to obtain their personalized perceptual similarity models. These personalized perceptual models are then used to obtain subject-specific similarity structure across the images in their encoding model training set. We fine-tune each participant's baseline encoder in two separate conditions, aligning its representations with either the generic or the personalized similarity judgments, with the same initialization and objective. We find that baseline brain encoders reflect within-subject perceptual similarity more strongly than measured fMRI responses and strongly preserve intersubject correlations present in measured fMRI, but fail to reflect the intersubject correlations of human perceptual similarity judgments. Fine-tuning brain encoding models with generic, population-level perceptual similarity information results in even better alignment with human perceptual judgments compared to baseline encoders, but these models still fail to capture intersubject correlation of human perceptual judgments. However, fine-tuning brain encoding models with personalized perceptual similarity information results in models that are overall the most aligned with human perceptual similarity judgments out of all the brain encoders, and also reflect intersubject correlation of human perceptual similarity judgments. Our framework provides a robust approach for bootstrapping human similarity judgment data into creating more individualized, human-aligned brain encoding models of vision.
\end{abstract}

\section{Introduction}

Visual brain encoding models aiming to predict brain responses elicited by images and video clips have become increasingly accurate as new vision models have been developed in adjacent fields. These visual brain encoding models are typically optimized to accurately predict measured brain activity ~\citep{mitchell_predicting_2008,huth_natural_2016,gu_personalized_2022,gifford_algonauts_2025,conwell_large-scale_2024,naselaris_encoding_2011}. However, a visual brain encoding model that accurately predicts measured brain activity may not necessarily have internal representations that reflect an individual person's perceptual similarity, i.e. how similar two images are to a particular person. Such alignment of brain activity models and human perceptual similarity would be particularly useful for downstream tasks - for example decoding, i.e. reconstructing seen or imagined stimuli from brain activity patterns. In decoding, having a model whose brain activity representations for any two images reflect those images' perceived similarity would allow reconstructing images with better fidelity to human similarity judgments~\citep{scotti_reconstructing_nodate, ozcelik_natural_2023, wang_mindbridge_2024, rajabi_human-aligned_nodate, takagi_improving_2023}.

\begin{figure}[!t]
\begin{center}
\includegraphics[width=0.9\linewidth]{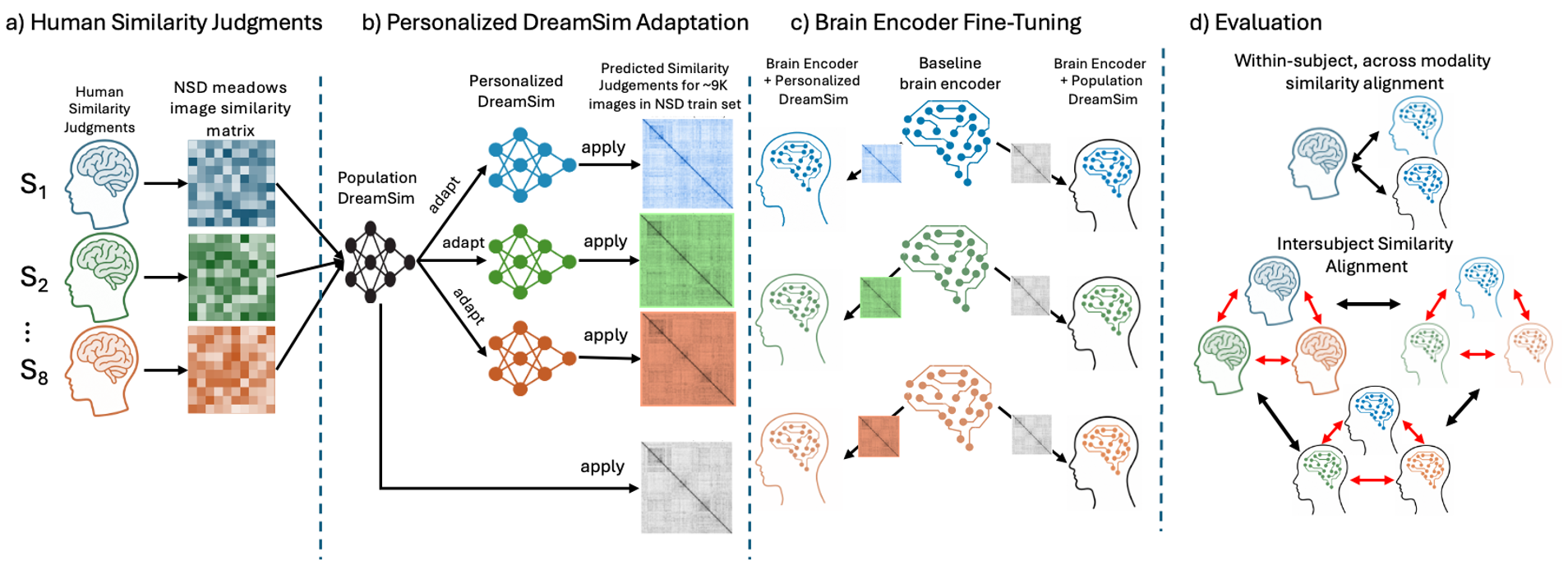}

%\framebox[4.0in]{$\;$}
%\fbox{\rule[-.5cm]{0cm}{4cm} \rule[-.5cm]{4cm}{0cm}}
\end{center}
\caption{Overview of the human image similarity judgment-brain encoding model alignment framework.
(a) Each of the eight NSD participants arranged an identical set of 100 images according
to their perceived similarity, producing a subject-specific human similarity judgment
matrix.
(b) A generic model of human perception (here called Population DreamSim) was adapted to each participant's observed similarity judgments to create
perceptual similarity models that reflect that person's judgments (here called personalized DreamSim). Population and personalized DreamSim models were used to
generate similarity values between each pair of images in a given participant's approximately 9,000 
training set images.
(c) Each participant's brain encoder was fine-tuned using both population
and personalized DreamSim estimates of image similarity to increase alignment of the brain encoder's predictions of brain responses with human perception.
(d) The resulting representations were evaluated by their alignment with human
similarity judgments and their preservation of across-subject similarity structure.}

\label{fig:framework}
\end{figure}

How a person perceives the similarity of two images, or human perceptual similarity, is a popular area of study in the broader field of psychophysics. Behavioral similarity experiments in these areas ask individuals to organize stimuli according to their perceived relationships. Representational similarity analysis (RSA) is a statistical approach that provides a framework for comparing similarity structures across various , multi-modal spaces~\citep{kriegeskorte_representational_2008, sucholutsky_getting_2024,nili_toolbox_2014, kriegeskorte_representational_2013, simkova_representations_2026}. RSA can be applied in this context to perceptual similarity, measured brain activity via fMRI, and encoding model predictions of brain activity to enable comparison of how image similarity spaces are conserved across the modalities. Analysis of these representational spaces can also allow quantification of either the similarity of representations across modalities within an individual or how the across-subject correlations of similarity spaces are preserved. A brain encoding model trained on individual participant data may be subject-specific in the sense that it predicts that participant's neural activity in an identifiable way (compared to other subjects' data), while still failing to reflect that person's perceptual relationships between stimuli.

Incorporating human similarity judgments into brain encoding models 
has several practical challenges.  Human similarity measurements
are onerous to collect for the thousands of image stimuli usually contained in high-quality vision neuroscience fMRI datasets \citep{allen_massive_2022,zerbe_laion-fmri_2026,hebart_things-data_2023,chang_bold5000_2019}. Therefore, when they are collected, it is only for a small subset of the images drawn from the stimuli set. Machine learning models have been used to estimate human similarity judgments, but are normally trained on data drawn across a cohort of many individuals and thus represent population-level averages rather than individual-level perceptual judgments \citep{marjieh_words_2023,fu_dreamsim_2023,muttenthaler_aligning_2025, zhang_unreasonable_2018}. These limitations motivate us to test if a small
amount of individual behavioral data can be used to personalize a scalable
perceptual model, and, subsequently, can the resulting estimated perceptual similarities be used to further personalize a brain encoder to increase alignment with that individual human's perception?

We investigate this question using the Natural
Scenes Dataset (NSD), which contains fMRI responses to tens of thousands of images from each of 8 participants, including a set of 1000 images that were seen by all participants - usually called the ``shared 1000'' \citep{allen_massive_2022}. Similarity judgments were collected from these 8 participants for a subset of 100 images drawn from the shared 1000 set. We begin by adapting DreamSim separately for each participant using a feature-reweighting function \citep{fu_dreamsim_2023} based on their similarity judgments, resulting in a Personalized DreamSim model for each participant (see Fig.~\ref{fig:framework}a--b). We then apply both the personalized DreamSim model and the population (original) DreamSim to the ~9,000 images in each of the 8 NSD participants' image training sets. We then use these two types of predicted similarities (from personalized or population-level DreamSim) to fine-tune subject-specific Neural Response Function (NRF) brain encoding models~\citep{chen_beyond_2026}. During fine-tuning, a
representational similarity objective encourages the encoder's predicted
voxel-response patterns to reflect either population-level or personalized
DreamSim perceptual similarity, while a frozen copy of the original encoder serves as a
teacher to constrain changes to its optimized brain activity predictions. This design allows
us to distinguish the effect of generic population-level perceptual alignment from the effect of individualized perceptual alignment. 

This work contains a few novel contributions, including
\begin{itemize}
    \item introduction of a framework for bootstrapping small amounts of subject-specific
    similarity judgments to adapt a pretrained, population-level perceptual similarity model to each individual.

    \item introduction of a method to fine-tune brain visual encoding models to increase alignment with both population-level and individual-specific human perception of images.

    \item an assessment of how increasing alignment of brain encoding models with human judgments changes their accuracy, and how these aligned models reflect within-subject perception and intersubject correlations in measured brain activity and perception of images.
    
\end{itemize}

\section{Related Work}

\paragraph{Visual brain encoding models and personalization.}
Visual brain encoding models aim to accurately predict neural responses elicited by
visual stimuli. \citep{naselaris_voxel-wise_2015,guclu_deep_2015,kay_identifying_2008,adeli_transformer_2026,st-yves_feature-weighted_2018}
Response-optimized encoding models have also been used to compare brain activity
representations to human ratings of visual similarity
 \citep{khosla_characterizing_2022}.
Personalized visual encoding models for a specific individual can preserve
inter-individual variability while reducing the amount of subject-specific
data required for accurate prediction
 \citep{gu_personalized_2022}.
These approaches personalize brain encoders according to fMRI-based measures of brain activity responses.
Our work asks a complementary question: \emph{can a brain encoder already tuned
to an individual's brain responses also be aligned with that same
individual's perceptual similarity structure?}

\paragraph{Human similarity judgment models and personalization.}
Human perceptual similarity judgments and measured brain responses may not possess
the same representational structure for identical visual stimuli
 \citep{king_similarity_2019}, and brain representations can contain
stable individual-specific structure related to differences in perceived
similarity
 \citep{charest_unique_2014}.
The work 
 \citet{brotherwood_revealing_2022} modeled subject-specific perceptual
structure using the same NSD dataset and showed that individual similarity structure 
can be recovered even with a relatively small 100-image task.
Our work takes this a step further by showing that observed subject-specific
similarity judgments are correlated with an individuals' brain activity and encoder, and, further, these observed judgments can be used to finetune a personalized both a perceptual judgment model and a brain encoder.

\paragraph{Human perception aligned computer vision models.}
Progress has been made in aligning computer vision model representations with human perception.
ML models designed to learn a perceptual similarity and
reflect human similarity judgments continue to improve 
as vision models improve \citep{fu_dreamsim_2023, muttenthaler_aligning_2025, roads_enriching_2021,peterson_evaluating_2018}.
Pretrained visual representations can also be transformed to improve their
correspondence with (population-level) human similarity judgments \citep{attarian_transforming_2021, muttenthaler_improving_2023},
and modern computer vision models can be fine-tuned using perceptual similarity
supervision \citep{sundaram_when_2024, zhao_shifting_2025}.
Because collecting human similarity judgments is costly, discovering ways to use a limited number of human judgments is fundamental
for making progress in this field. 
 \citep{marjieh_words_2023,sucholutsky_informativeness_2023}.
 \citet{muttenthaler_aligning_2025} uses limited (population-level) human similarity judgments to
construct human-aligned supervision that can be generated over much larger
image sets and transferred into vision foundation models.
Our framework applies this general principle of alignment of human perception for the first time to brain encoders, doing so at an individual subject level rather than using population-level perceptual data.
\section{Methods}

\subsection{Data, Human Similarity Judgments, and Baseline Brain Encoders}

In this paper we used the Natural Scenes Dataset (NSD) \citep{allen_massive_2022} that contains whole-brain 7T fMRI measurements of activity responses in 8 individuals while viewing tens of thousands of natural
scene images drawn from MS COCO \citep{lin_microsoft_2015}. FMRI data was processed and brain responses were estimated using
GLMsingle \citep{prince_improving_2022}, normalized within each session, and averaged
across repeated presentations of the same image. Each subject viewed approximately 9,000 unique images in addition to 1,000 images shared across every subject. We apply the official nsdgeneral region-of-interest (ROI) mask to this data to work only with the voxels found in the 7 lower-level and 12 higher-level visual cortical regions common to all participants. 

Individual perceptual similarity judgments were obtained from participants using the NSD-Meadows multiple-arrangement task \citep{kriegeskorte_inverse_2012}. In this task, the 8 NSD
participants evaluated the same set of 100 images, drawn from the shared 1,000 images. They were asked to arrange the images inside a circle, placing more similar images nearer to one another; they were given 45 minutes to complete the task. In the end, the distance between the circle positions was calculated to obtain image-pair distances, yielding a $100\times100$ matrix. We negated the distance in order to have a metric that represented human similarity, where larger values indicate greater perceived similarity.
We use pretrained subject-specific Neural Response Function (NRF) models as the baseline brain encoding models. An NRF model was trained for each individual in their unique ~9K image response pairs to predict voxel-wise fMRI responses jointly from visual features and voxel location \citep{chen_beyond_2026}. All subsequent encoding model fine-tuning initialize training with these individual-specific baseline brain encoding models.

\subsection{Personalized Human Similarity Judgment Models}

Measured human similarity judgments are available for only 100 images per participant, while each brain encoder is trained on approximately 9,000 images. If we are to fine-tune our encoding models to better align them with an individual's similarity judgments, it would be better if we had similarity judgments for all 9,000 training images. As a result, we adapt DreamSim separately to each participant in order to obtain subject specific similarity predictions for all pairs of images in the entire 9,000 unique image set. We begin with the pretrained generic (population-level) DreamSim model, which for each image produces a 1,792-dimensional image embedding vector in a space that was optimized to reproduce population-level human similarity judgments, i.e. vectors closer to one another in this space are perceived as more similar by people. During personalization, the DreamSim model remains
frozen, but for each $k$ subject we learn a feature-reweighting
vector $\mathbf{w}_k\in\mathbb{R}^{1792}$ to optimize the match to that particular person's similarity judgments. Specifically, for images $i$ and $j$,
the personalized similarity judgment model $S$ for subject $k$ is

\begin{equation}
    S_k(i,j)
    =
    \cos\left(
        \mathbf{w}_k \odot \mathbf{e}_i,\,
        \mathbf{w}_k \odot \mathbf{e}_j
    \right),
\end{equation}

where $\mathbf{e}_i$ and $\mathbf{e}_j$ are pretrained DreamSim embeddings for images $i$ and $j$ and $\odot$ indicates element-wise multiplication of the two vectors. The vector $w_k$ is optimized to match the participant's observed similarity judgments $S_k^{obs}$:

\begin{equation}
\begin{aligned}
    \mathcal{L}^k_{\mathrm{adapt}}
    &=
    \frac{1}{|\mathcal{P}|}
    \sum_{(i,j)\in\mathcal{P}}
    \left[
        z(S_k(i,j))
        -
        z(S_k^{\mathrm{obs}}(i,j))
    \right]^2
    +
    \lambda
    \frac{1}{D}
    \left\|\mathbf{w}_k-\mathbf{1}\right\|_2^2 ,
\end{aligned}
\end{equation}

with $D=1792$, $\lambda=1$, $\mathcal{P}$ is the set of all pairs of images, and $z$ indicates we normalize (via z-score) the model's predicted similarities and observed similarities before taking their difference. The second term in the loss function regularizes $w_k$ to not deviate too far from the unit vector that represents the original DreamSim model output. Optimization uses full-batch Adam
($\mathrm{lr}=0.05$) for 100 steps. We optimize $\mathcal{L}_{\mathrm{adapt}}$  using repeated 5-fold cross-validation. In each repeat, the 100 Meadows images are divided into five folds of 20 images. DreamSim is adapted using 4 of the 5 folds (80 images or 3,160 image pairs) and evaluated only on the 190 image pairs formed by the 20 held-out images; there is no crossover of images in the training and testing sets. We repeated the cross-validation procedure 10 times using independent random fold assignments and averaged the resulting held-out performance estimates across repetitions. In the end, we have two human judgment models: a Population DreamSim, which is the original, generic model, and Personalized DreamSim models that are characterized by their unique re-weighting vector for each subject. 

\subsection{Aligning Brain Encoders to Human Similarity Judgment Models}

For each subject, Population and Personalized DreamSim are applied to the approximately 9,000
images used to train that person's Baseline Brain Encoder, producing an estimated human similarity judgment vector of size $\binom{9,000}{2}=4.04\times10^{7}$ for each of the two DreamSim models, denoted $\mathbf{S}_k^{\mathrm{DS}}$. A student brain encoder is initialized from the participant's Baseline Brain Encoder, while a frozen copy of the baseline model serves as a teacher. The student encoder of Subject $k$ produces predicted voxel-response patterns for the set of training images, which are then correlated (via Pearson) across pairs of images and collected in $\mathbf{S}_k^{\mathrm{student}}$. Fine-tuning aims to jointly preserve the teacher's (Baseline Brain Encoder's) voxel-level activation predictions and align the student encoding model's representational similarity structure $\mathbf{S}_k^{\mathrm{student}}$ with either the population level or personalized DreamSim estimated human similarity structure $\mathbf{S}_k^{\mathrm{DS}}$:

\begin{equation}
    \mathcal{L}^k_{\mathrm{finetune}}
    =
    \lambda_{\mathrm{teacher}}
    \mathrm{MSE}\left(
        \hat{\mathbf{Y}}_k^{\mathrm{teacher}},
        \hat{\mathbf{Y}}_k^{\mathrm{student}}
    \right)
    +
    \lambda_{\mathrm{RSA}}
    \left[
        1-r\left(
            \mathbf{S}_k^{\mathrm{student}},
            \mathbf{S}_k^{\mathrm{DS}}
        \right)
    \right],
\end{equation}

where $\hat{\mathbf{Y}}_k^{\mathrm{teacher}}$ and $\hat{\mathbf{Y}}_k^{\mathrm{student}}$ are the voxelwise brain activiy predictions from the two encoding models, $\lambda_{\mathrm{teacher}}=10$, and $\lambda_{\mathrm{RSA}}=2$. With this loss function, we fine-tune two brain encoders. Brain Encoder + Population DreamSim refers to the brain encoder fine-tuned with similarity judgments from the population DreamSim, and Brain Encoder + Personalized DreamSim refers to the brain encoder fine tuned with similarity judgments from personalized DreamSim.  Models are optimized using Adam ($\mathrm{lr}=10^{-4}$) for up to 15 epochs across five folds and 10 different 5-fold cross-validation assignments for all 8 subjects. Predictions from the
five held-out folds are combined within each repeat, and fold-dependent
quantities are then averaged across the 10 repeats.

\subsection{Model Evaluation and Statistical Analysis}

For each participant and representational source (measured fMRI brain responses, human similarity judgments, or brain encoding models), we construct a
representational similarity matrix (RSM), where larger values indicate greater image similarity. Brain-encoder and measured fMRI RSMs are defined using Pearson correlation between predicted or measured voxel-response patterns, calculated using the mask of 19 visual ROIs present across all 8 individuals in the NSD. A view of the 19 ROIs used is provided in 
Appendix Fig.~\ref{fig:roi}. DreamSim RSMs use cosine similarity, and human RSMs use the negated pairwise dissimilarity estimates derived from the NSD-Meadows Multiple Arrangements Task.

Within-subject human similarity alignment is measured as Spearman correlation between a data- or model-derived RSM and the corresponding participant's human similarity judgment RSM. Paired model comparisons of within-subject human similarity alignment use two-sided Wilcoxon signed-rank tests. To quantify
intersubject structure, we compute Spearman correlation between RSMs
of each pair of participants across different modalities, yielding an $8\times8$ intersubject similarity
matrix for each representational source that reflects how similar two people's RSMs are across modalities. Within-modality, across subject similarities were calculated, and their intersubject correlational spaces compared across modalities using Spearman rank correlation to test if across-subject relationships in RSMs were preserved across modalities. Correlations between intersubject similarity matrices across various modality pairs are evaluated using a Mantel-style subject-label permutation test with 10,000 permutations to account for dependence among the 28 subject pairs. Fold-dependent quantities are averaged across the 10 cross-validation repeats before statistical testing.
\section{Results}
\label{sec:results}

\subsection{Within and across subject comparisons of multi-modal representational similarity matrices}

\begin{figure}[!h]
\begin{center}
\includegraphics[width=0.8\linewidth]{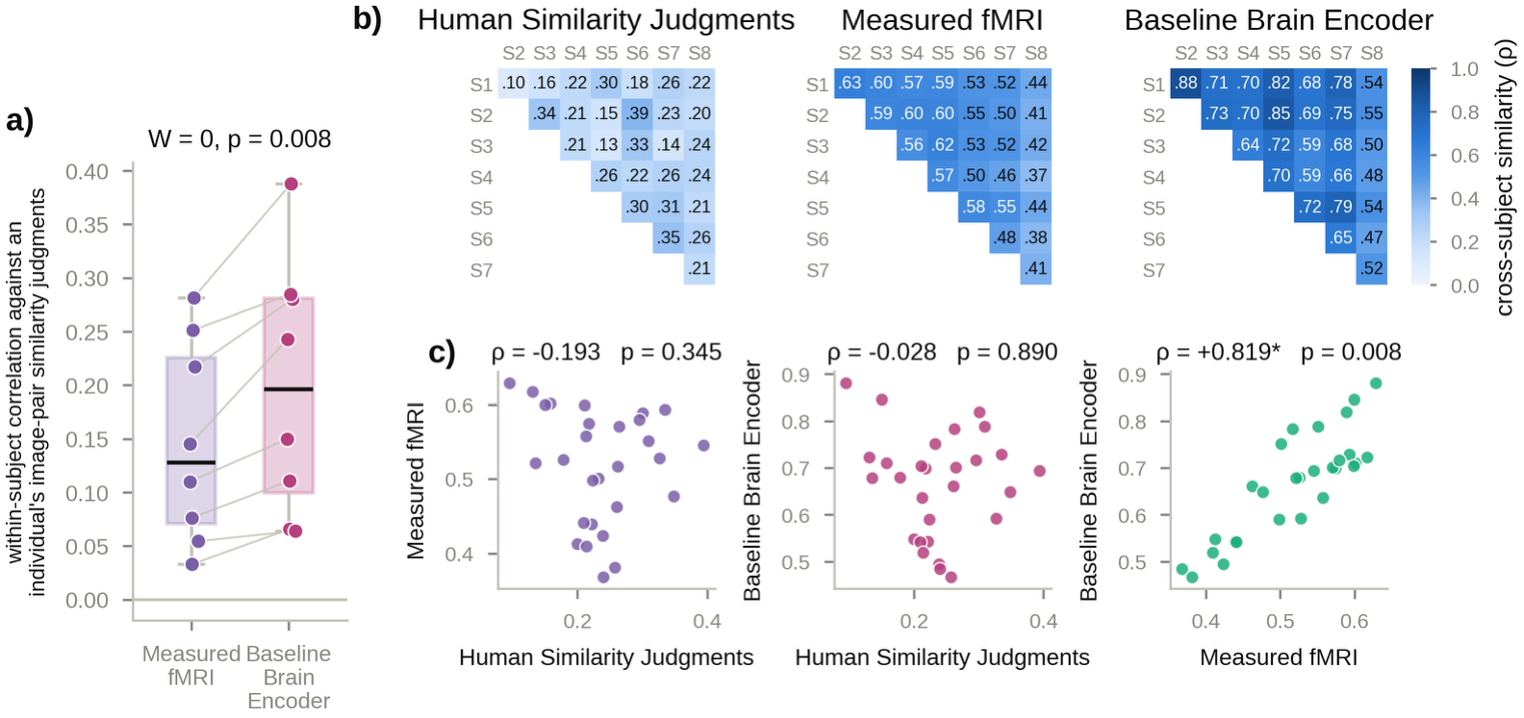}
\end{center}
\caption{(a) Within subject alignment of an individual's perceptual similarity judgments for pairs of images in the set of 100 images used in the perceptual task and i) measured fMRI (left, purple) and ii) baseline brain encoder predictions(right, pink) for the same pairs of images.
(b) Intersubject similarity matrices derived from human similarity judgments,
measured fMRI, and Baseline Brain Encoder predictions for image-pairs in the set of 100 images across the eight NSD subjects. (c) Spearman correlation between the 28 unique subject-pair values
of each intersubject similarity matrix across the modalities in panel b). Intersubject baseline encoder structure strongly
matches measured fMRI ($\rho=0.82$, $p=0.008$), but neither measured fMRI nor
the baseline encoder matches behavioral intersubject similarity judgment structure
($\rho=-0.19$ and $-0.03$, both $p>0.05$).
}

\label{fig:baseline}
\end{figure}

We begin by computing the 8 individuals' representational similarity matrices for 1) human similarity judgments, 2) measured fMRI brain responses, and 3) baseline brain encoders, using the 100 images contained in the NSD's perceptual similarity task. We wanted to test if images that were rated as more similar by a specific individual had more similar measured or predicted brain activity. To do this, we correlated via Spearman the RSMs for human judgment with both the i) measured fMRI and ii) baseline brain encoder. Fig.~\ref{fig:baseline}a  shows the 8 subjects' correlations of these RSMs. In all individuals, both the correlations between the RSMs for human judgment and measured fMRI brain activity were significant. However, interestingly, the baseline brain encoding model had significantly better alignment ($W = 0$, $p=0.008$) with human similarity judgments ($\rho =0.198$, all 8 subjects $p<0.001$) compared to the measured fMRI brain response alignment ($\rho =0.146$, all 8 subjects $p\leq0.018$). We posit that this may be due to the boost in SNR that the encoding model has over the raw fMRI data.

We then test if the intersubject correlations of multi-modal RSMs were preserved across modalities, i.e. if two people had more similar perceptual similarity judgments did they also have more similar measured and/or predicted brain activity. We characterized intersubject structure by correlating the three RSMs (measured fMRI, baseline brain encoder, and human similarity judgments) for each pair of subjects, see  Fig.~\ref{fig:baseline}b. As expected, we found that the intersubject structure of the Baseline Brain Encoders closely matches the structure
of measured fMRI ($\rho=0.819$, $p=0.008$), see Fig.~\ref{fig:baseline}c. In contrast,
human similarity judgments' intersubject structure was not significantly correlated with either
measured fMRI ($\rho=-0.193$, $p=0.345$) or the Baseline Brain Encoders
($\rho=-0.028$, $p=0.890$), see Fig.~\ref{fig:baseline}c. We conclude that although the subject specific Baseline Brain Encoders
preserve individual differences present in measured brain responses, these differences
do not correspond to the across-individual structure that exists in perceptual
similarity judgments.

\subsection{Personalized DreamSim models and Brain Encoder Finetuning}

The population level DreamSim predictions had moderate correlations with each of the 8 individuals' measured similarity judgments ($\rho = 0.245 \pm 0.030$ across subjects , every individual correlation was significant at $p<0.001$). Our next goal was to adapt the population level DreamSim using subject-level similarity judgments to build perceptual models that better represented specific individuals. We found that the personalized DreamSim did have a statistically significant increase in alignment to the actual individual's similarity judgments ($\rho = 0.314 \pm 0.025$, all $p<0.001$) compared to the Population DreamSim (mean gain +0.069, higher in 8/8 subjects, $W = 0$ $p=0.008$). 
Population DreamSim and Personalized DreamSim models were then used as proxies to create estimates of individuals' similarity matrices for all of the image pairs in their baseline brain encoder training set, which was then used for the next step described below - fine-tuning the baseline brain encoders.

\begin{figure}[!h]
\centering
\includegraphics[width=0.8\linewidth]{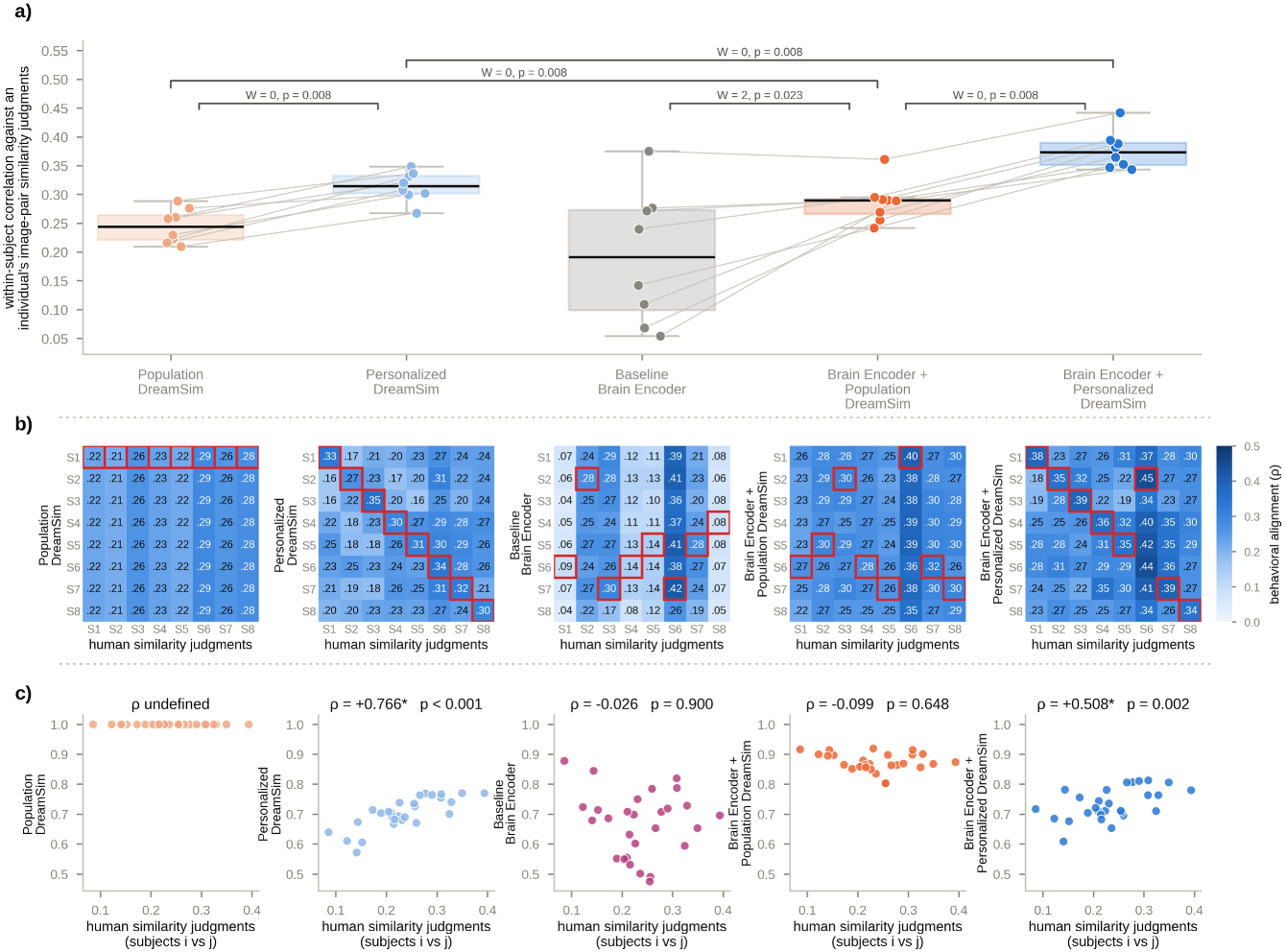}
\caption{(a) Within-subject alignment of an individuals' image-pair similarity judgments for the 100 images in the perceptual task compared to the representational similarity of: population DreamSim, personalized DreamSim, baseline brain encoder, brain encoder + population DreamSim, and brainencoder + personalized DreamSim, for the same image-pairs. Bars above pairs of boxplots indicate those with significant differences ($p<0.05$).
(b) Human similarity judgment alignment (Spearman $\rho$) between each subject-specific model
(rows) and each subject's human similarity judgments (columns); red boxes mark
the maximum value for each column. Boxes on the diagonal indicate that person's model best matched to their observed similarity judgments compared to others' similarity judgments.
(c) Intersubject correlations of representational spaces for human similarity judgments vs intersubject correlations of representational spaces the same modalities listed in panel (b). Each of the 28 points represents a unique pair of individuals, where the x-axis is the correlation of those two individuals' human similarity judgment RSMs and the y-axis is the correlation of those two individuals' RSMs for that specific modality.}
\label{fig:personalized}
\end{figure}
We compared three brain encoders: the baseline brain encoder, the baseline brain encoder finetuned with Population DreamSim (Brain Encoder + Population DreamSim), and the baseline brain encoder finetuned with Personalized DreamSim (Brain Encoder + Personalized DreamSim). Within subject alignment of human judgment RSMs and brain encoder RSMs increased significantly across the three models: Baseline Brain Encoder (mean $\rho=0.192$, all $p\leq 0.032$), Brain Encoder + Population DreamSim (mean $\rho=0.287$, all $p<0.001$), and Brain Encoder + Personalized DreamSim (mean $\rho=0.377$, all $p<0.001$), see Fig.~\ref{fig:personalized}a. The Baseline Brain Encoder RSMs had significantly lower correlation with human similarity judgment RSMs compared to the Brain Encoder + Population DreamSim ($W=2$, $p=0.023$) and the Brain Encoder + Personalized DreamSim ($W = 0$, $p=0.008$), the latter two also being significantly different ($W=0$, $p=0.008$). The Brain encoder + personalized DreamSim also had higher correlations to human similarity judgments within individuals compared to just Personalized DreamSim (mean $\rho = 0.377$ vs $0.314$ $W=0$, $p=0.008$).

\subsection{Multi-modal Intersubject Correlation Structure}

Improved within-subject, across image-pair human similarity judgment alignment of encoding models does not mean that these models capture
the across subject relationships of the various modalities. We therefore asked
whether the intersubject organization of each model corresponded to the
intersubject structure observed in human similarity judgments.

As expected, Personalized DreamSim reproduced the intersubject structure of the
human similarity judgments ($\rho=0.766$, $p<0.001$). Brain Encoder +
Personalized DreamSim also recovered this structure
($\rho=0.508$, $p=0.002$). Neither the Baseline Brain Encoder
($\rho=-0.026$) nor Brain Encoder + Population DreamSim
($\rho=-0.099$) showed significant correspondence with
intersubject correlations of human similarity judgments, see Fig.~\ref{fig:personalized}c. The personalized-encoder correlation was positive
across all 10 cross-validation repeats (range across repeats was $\rho=0.38$--$0.57$),
showing that this effect was robust to the assignment of images
to cross-validation test folds. The comparison with measured fMRI showed a different pattern.
The Baseline Brain Encoder strongly preserved measured-fMRI intersubject
human similarity judgment structure ($\rho=0.84$, $p=0.007$) as expected, and this correlation remained high following
population-level alignment ($\rho=0.77$, $p=0.014$). After personalized alignment,
correlation with measured-fMRI intersubject structure was not significant.

These results show a contrast in the effect of population perceptual finetuning vs
personalized perceptual finetuning on brain encoders. Population DreamSim finetuning reshaped the brain encoder to be better aligned with human similarity judgments overall (compared to the baseline brain encoder) but did not reflect across-subject similarity judgment correlations; further, this model still reflected intersubject correlations of measured fMRI responses. In contrast, Personalized DreamSim finetuning reshaped the brain encoder to better reflect within-subject human similarity judgments as well as across-subject similarity judgment correlations, but no longer reflected intersubject similarity of measured fMRI data.
\begin{figure}[!t]
\begin{center}
\includegraphics[width=0.8\linewidth]{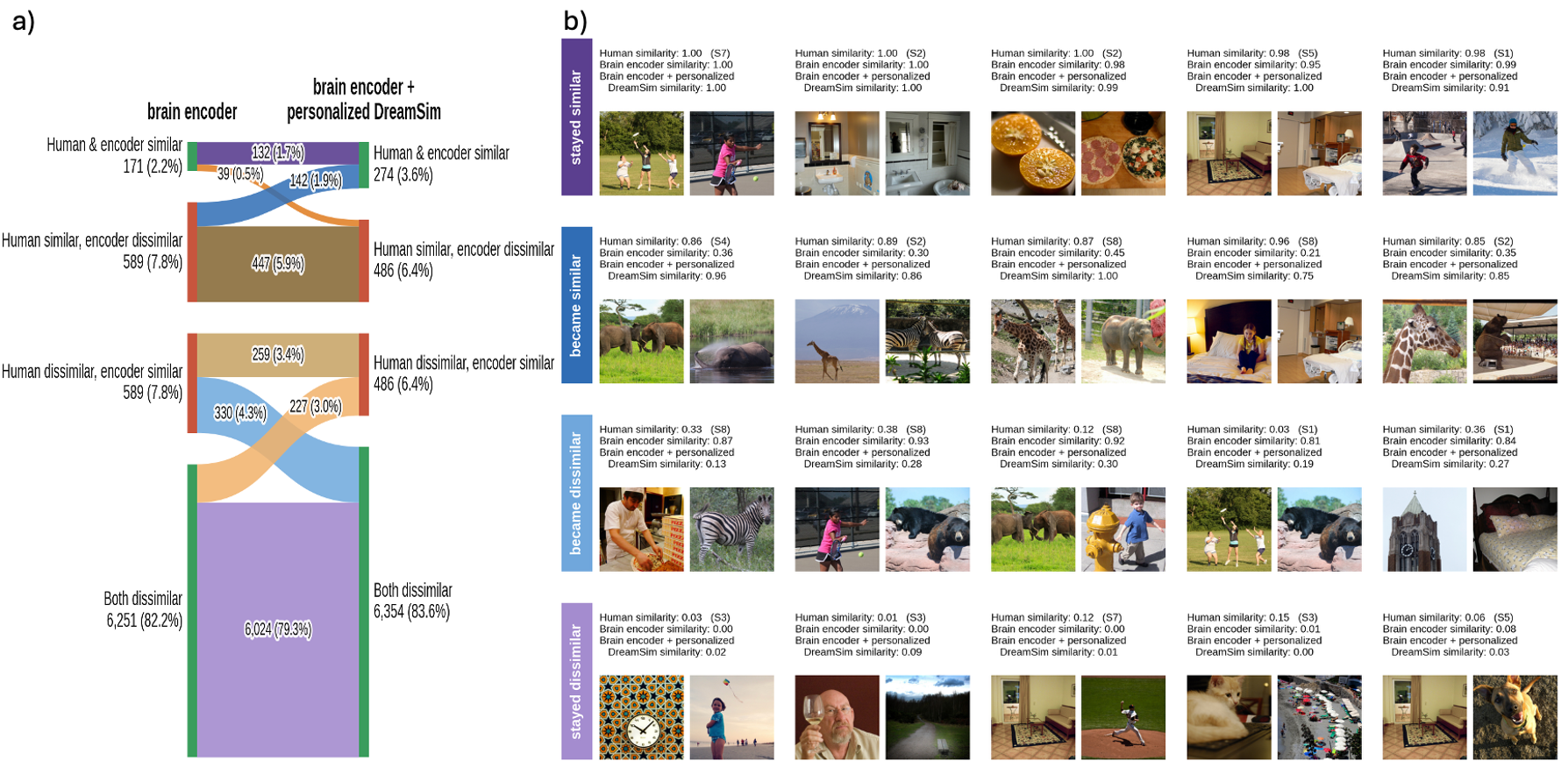}
\end{center}
\caption{
(a) A Sankey plot showing changes in image-pair similarity agreement between human judgments and the
brain encoder before (left column) and after (right column) Personalized DreamSim fine-tuning. Red vertical bars indicate humans and encoders did not agree in their similarity assessments for that image-pair, while green vertical bars indicate the two models did agree. 
(b) Representative image-pairs illustrating the different group agreement transition types (from baseline brain encoder to brain encoder + Personalized DreamSim), where the color behind each vertical label matches the image-pair group in panel (a). Image-pair similarity scores for a specific human's judgments (i.e. ``S7'' indicates subject 7), baseline brain encoder, and brain encoder + Personalized DreamSim are above each image pair.
}
\label{fig:sankey}
\end{figure}

\subsection{Brain Encoder Accuracy vs. Human Judgment Alignment: a Tradeoff}

To ensure we are not losing too much brain activity prediction accuracy during the brain encoder finetuning, we also examined how human similarity judgment alignment affected voxel-level prediction accuracy. For each subject, we compared the finetuning procedure's impact on both 
behavioral alignment and voxel-level prediction accuracy, see Appendix Fig.~\ref{fig:tradeoff}. Some subjects showed large gains in behavioral alignment with
relatively modest changes in neural-prediction accuracy, and others showed
larger reductions in voxel-prediction performance. Overall, however, the relative change in voxel accuracy was modest (mean -9.2\% for the personalized encoder, -7.1\% for the population one) compared to the gains in human similarity alignment (mean +190\% and +119\% respectively). Human similarity alignment of brain encoders also generally increased with finetuning (baseline $<$ population DreamSim $<$ Personalized DreamSim), while
fine-tuning related changes in voxel-prediction accuracy varied substantially across
participants. ROI-wise analyses further characterize how perceptual alignment is distributed
across the visual hierarchy for measured fMRI and each brain-encoder condition (Appendix Fig.~\ref{fig:roi}).
These results demonstrate a subject-dependent accuracy vs alignment tradeoff, with relative gains in human similarity alignment being much larger than the relative losses in brain encoder prediction accuracy. 

\subsection{Image-Pair Changes}

Finally, we examined how individual
image pairs changed in their similarity spaces across the various modalities, see Fig.~\ref{fig:sankey}a. In the figure, red vertical lines indicate the set of image pairs that disagreed in their similarity assessment among the human judgments and the brain encoder, while green vertical bars indicate the set of image pairs that agreed in their similarity assessment. The first column (having longer red bars and shorter green bars) represents the baseline brain encoder and the second column represents the brain encoder + Personalized DreamSim. 

Out of the possible shifts in agreement groups, only 3.5\% of image pairs moved from human similarity judgment and brain encoder agreement to a disagreement when going from the baseline brain encoder to the personalized DreamSim brain encoder. In contrast, almost double (6.2\%) the number of image pairs moved from a disagreement state to an agreement state. Example image pairs for the various agreement/disagreement groups provided in Fig.~\ref{fig:sankey}b illustrate the
types of perceptual relationships altered during fine-tuning. This analysis illustrates the benefit of individualized, perceptual fine-tuning in brain encoding models to increase alignment of specific image pairs to better reflect a given person's perception. 

%\input{figures/figure5}

%Among image pairs judged similar by participants but marked
%as dissimilar by the Baseline Brain Encoder (7.8\% of image pairs), approximately %26\% of them became similar
%following personalized fine-tuning. On the other hand,  out of the
%image pairs determined to be dissimilar by humans and similar by the baseline
%encoder (7.8\% of image pairs), 61\% became dissimilar in the brain encoder space %following fine-tuning.

%These corrections were accompanied by comparatively smaller opposing
%transitions to image pairs with disagreements between the human similairy %jugements and brain encoder. Of the 2.2\% of image pairs with matching similarity %assessments in the baseline brain encoder, only 22.5\% became dissimilar, while  %3.9\% of previously matched
%human-dissimilar pairs became similar. 

\section{Discussion and Conclusion}

Our results show that baseline brain encoders can be highly subject-specific in their predictions while still failing to reflect how that individual perceives relationships between visual stimuli.  Population-level perceptual
fine-tuning improved brain encoders' perceptual alignment, yet still failed to recover
individually-specific perception. Personalized perceptual models were able to 
capture both within-subject perceptual judgments and intersubject similarity judgment
structure, in turn transferring these properties to the 
personalized brain encoders. 

Our approach is limited by DreamSim's performance as a proxy for human judgment. With continued advances in vision and vision-language foundation models,
modernizing DreamSim using newer perceptual representations could provide a
stronger population-level similarity baseline. Future directions for this work can focus on evaluating whether different perceptual models can act as stronger generalizable targets for personalized brain encoder fine-tuning. 

These findings establish a framework for incorporating a small number of human similarity judgment
measurements into brain encoding models to make them more aligned to an individual's perception. Doing so moves visual brain encoding models toward internal representations that reflect not only how a person's brain activates in response to an image, but also that person's perception of image similarity.

\bibliography{iclr2027_conference}
\bibliographystyle{iclr2027_conference}

\subsection*{AI use statement}
In this work, we used generative AI tools for aiding with writing help in checking for spelling errors, grammar mistakes, and finding new word choices.We also use generative AI tools for assisting with coding tasks such as creating figures. LLM generated code was verified and tested for correctness.  We take responsibility for the final content of this work,
including text, claims or artifacts produced with the aid of generative AI.

\subsection*{Ethics Statement}
Our research adheres to the ICLR Code of Ethics. All experiments in this paper are conducted using
open-source datasets, and no potential ethical concerns are associated with this work.

\subsection*{Reproducibility Statement}
All preprocessed data, code, and model parameters used in our research will be made publicly
available upon publication. Detailed protocols for data preprocessing, model training, and evaluation
have been provided in our manuscript, enabling independent reproduction

\appendix
\section{Additional Results}
\label{Appendix A}
\begin{figure}[!h]
\begin{center}
\includegraphics[width=0.75\linewidth]{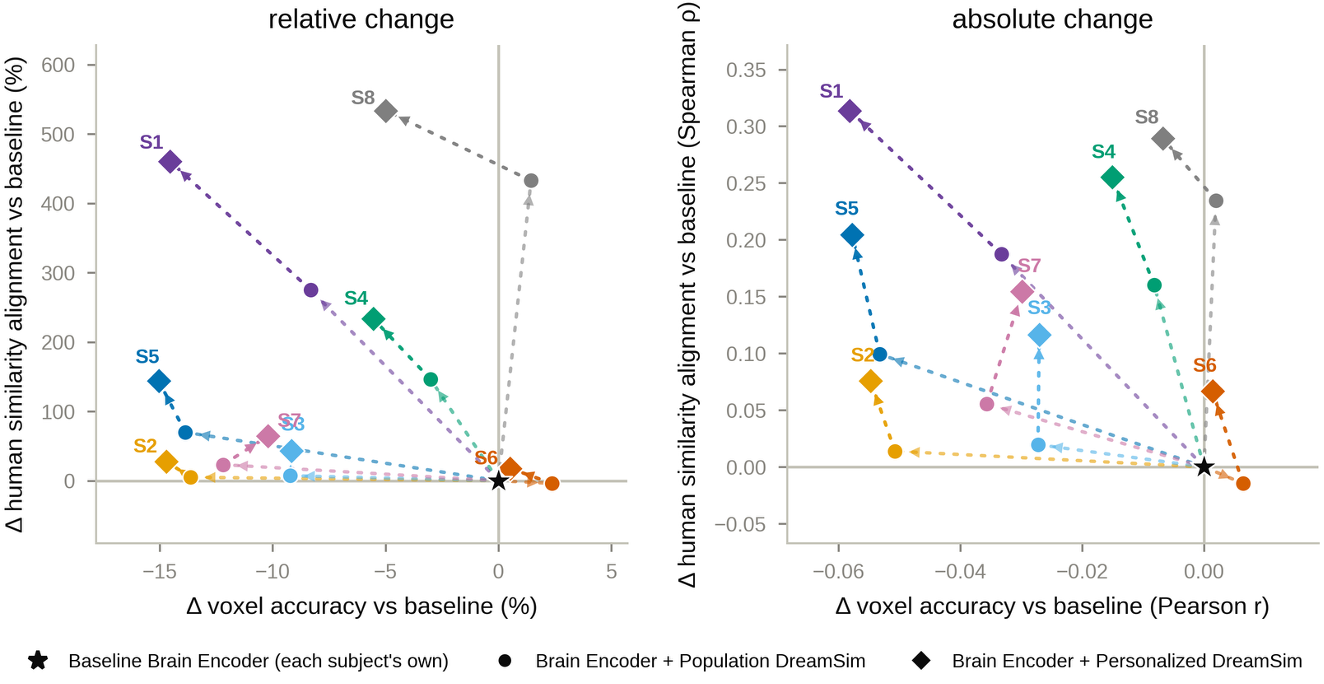}
\end{center}

\caption{
Subject-specific relative change in brain encoder accuracy plotted against that same individuals' change in human similarity judgment alignment, with each subject a different color. Baseline Brain Encoders are marked as stars, Brain Encoder + Population DreamSim are marked as circles, and Brain Encoder + Personalized DreamSim are marked as diamonds. Dashed lines with arrows emphasize the flow from Baseline Brain Encoder $\rightarrow$ Brain Encoder + Population DreamSim $\rightarrow$ Brain Encoder + Personalized DreamSim. 
}
\label{fig:tradeoff}

\end{figure}
\begin{figure}[!h]
\begin{center}
\includegraphics[width=0.9\linewidth]{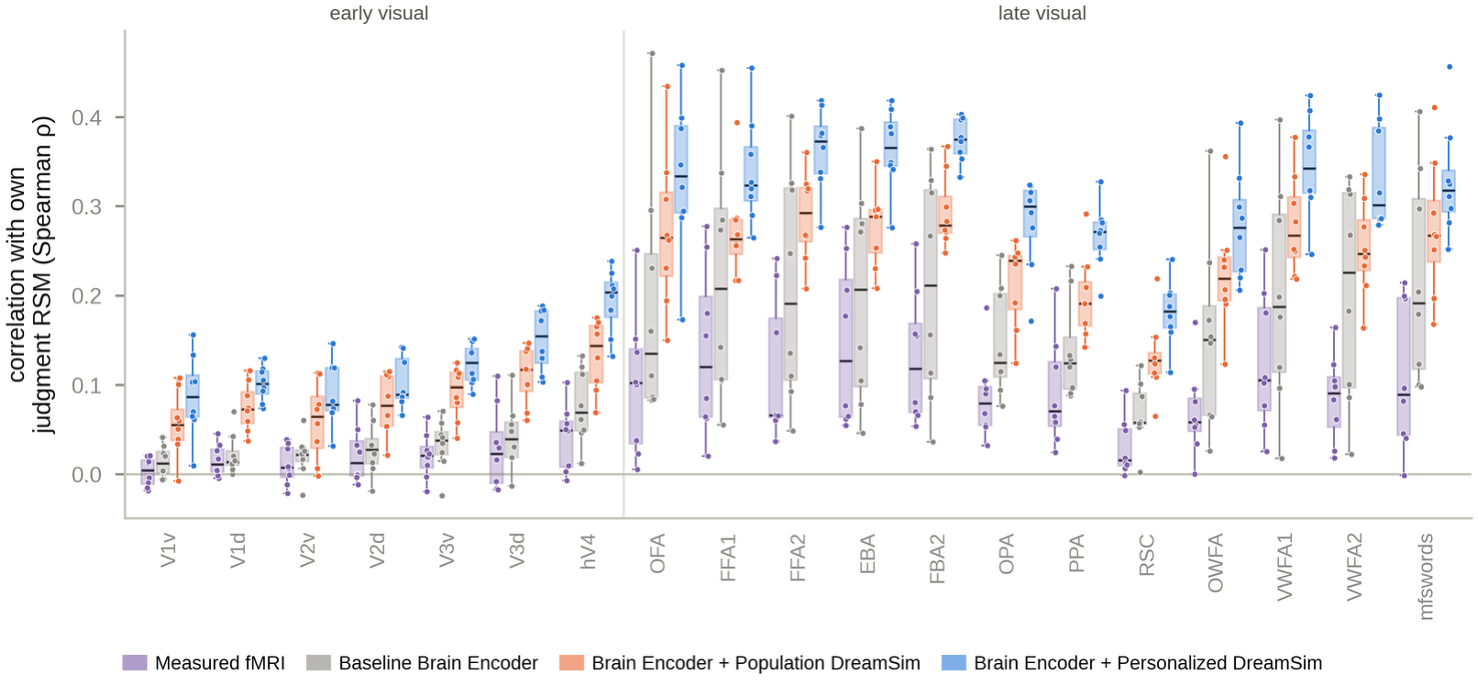}
\end{center}
\caption{
Region-wise correlation between brain and human representational similarity matrices, for the 19 ROIs common to all eight subjects and ordered early $\rightarrow$ late along the visual hierarchy. Each region's RSM is built from that region's voxels alone and correlated with that same individual's human similarity judgment RSM; boxes summarize the eight subjects and the overlaid points are those same eight values. Measured fMRI is shown in purple, Baseline Brain Encoders in gray, Brain Encoder + Population DreamSim in orange, and Brain Encoder + Personalized DreamSim in blue. All four are scored on the pooled held-out pairs of 5-fold cross-validation (950 pairs per subject), averaged over ten fold assignments.
}
\label{fig:roi}
\end{figure}

\end{document}